\documentclass[twocolumn,twoside,slac_two]{revtex4}
\usepackage{graphicx}
\usepackage{fancyhdr}
\begin{document}

\title{Modeling High-energy and Very-high-energy $\gamma$-rays from the Terzan~5 Cluster}

%
\author{C. Venter}
\affiliation{Centre for Space Research, North-West University, Potchefstroom Campus, Private Bag X6001, Potchefstroom 2520, South Africa}

\author{O.C. de Jager}
\affiliation{Centre for Space Research, North-West University, Potchefstroom Campus, Private Bag X6001, Potchefstroom 2520, South Africa}
	     
\author{Andreas Kopp}
\affiliation{Institut f\"{u}r Experimentelle und Angewandte Physik, Christian-Albrechts-Universit\"{a}t zu Kiel, Leibnizstrasse 11, 24118 Kiel, Germany}

\author{I. B\"usching}
\affiliation{Institut f\"{u}r Theoretische Physik, Lehrstuhl IV: Weltraum- und Astrophysik, Ruhr-Universit\"{a}t Bochum, 44780 Bochum, Germany}

\author{A.-C. Clapson}
\affiliation{Max-Planck-Institut f\"{u}r Kernphysik, PO Box 103980, 69029 Heidelberg, Germany}

\begin{abstract}
The \textit{Fermi} Large Area Telescope (LAT) has recently detected a population of globular clusters (GCs) in high-energy (HE) $\gamma$-rays. Their spectral properties and energetics are consistent with cumulative emission from a population of millisecond pulsars (MSPs) hosted by these clusters. For example, the HE spectra exhibit fairly hard power-law indices and cutoffs around a few GeV, typical of pulsed spectra measured for the $\gamma$-ray pulsar population. The energetics may be used to constrain the number of visible MSPs in the cluster ($N_{\rm vis}$), assuming canonical values for the average $\gamma$-ray efficiency and spin-down power. This interpretation is indeed strengthened by the fact that the first $\gamma$-ray MSP has now been identified in the GC NGC~6624, and this MSP is responsible for almost all of the HE emission from this cluster~\cite{Parent11}. On the other hand, it has been argued that the MSPs are also sources of relativistic leptons which may be reaccelerated in shocks originating in collisions of stellar winds in the cluster core, and may upscatter bright starlight and cosmic microwave background photons to very high energies. Therefore, this unpulsed component may give an independent constraint on the total number of MSPs ($N_{\rm tot}$) hosted in the GC, for a given cluster magnetic field $B$ and diffusion coefficient $k_0$. Lastly, the transport properties of the energetic leptons may be further constrained using multiwavelength data, e.g., to infer the radial dependence of $k_0$ and $B$. We present results on our modeling of the pulsed and unpulsed $\gamma$-ray fluxes from the GC Terzan~5.
\end{abstract}

\maketitle

\thispagestyle{fancy}


\section{INTRODUCTION}
\begin{figure}[t]
\centering
\includegraphics[width=80mm]{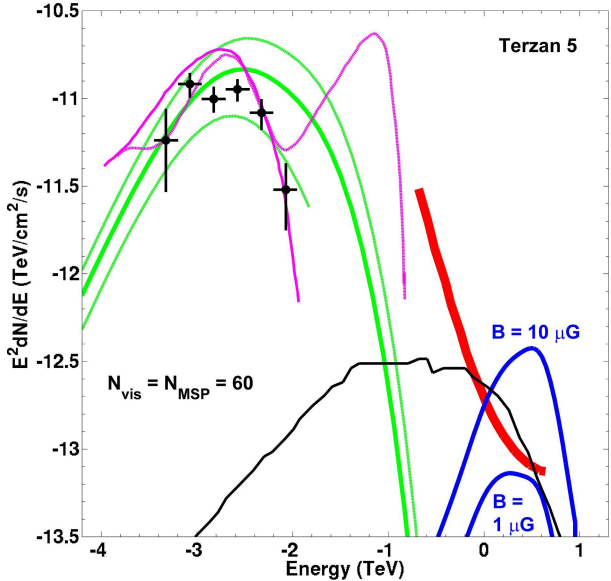}
\caption{Differential CR and IC spectra for Terzan~5. The green lines are the CR spectrum and errors from~\cite{Venter09_GC}, scaled using $N_{\rm vis}=60$ to fit the \textit{Fermi} data; the magenta lines are IC models from~\cite{Cheng10}; \textit{Fermi} LAT data~\cite{Abdo10} are plotted in black; the black line is an IC prediction from~\cite{BS07}, scaled to the 
new distance and bolemetric luminosity, average spin-down luminosity of $1.8\times10^{34}$~erg~s$^{-1}$, and $N_{\rm tot}=60$; the blue lines represent IC spectra for different values of $B$ (this work); the red line is the H.E.S.S.\ sensitivity.} 
\label{fig1}
\end{figure}
The recent \textit{Fermi} LAT detection of several globular clusters (GCs) in high-energy (HE) gamma rays~\cite{Abdo09,Abdo10,Kong10,Tam11}, very plausibly including Terzan~5, underlined the importance of modeling collective $\gamma$-ray emission of millisecond pulsars (MSPs) in GCs. The HE spectra are thought to represent the cumulative contribution of magnetospheric radiation from a population of MSPs hosted by the GC. This spectral component has been calculated for 47~Tucanae~\cite{Venter08} and Terzan~5~\cite{Venter09_GC} in the context of curvature radiation (CR) by primary electrons being constrained to move along curved magnetic field lines in the magnetospheres of an ensemble of MSPs.

An alternative calculation~\cite{Cheng10} considered a scenario where inverse Compton (IC) scattering, and not CR, was responsible for the HE fluxes seen by \textit{Fermi}. The \textit{Fermi} fluxes may be reproduced for certain parameters, and this model also predicted very-high-energy (VHE) components in some cases. The discovery of the luminous MSP PSR~J1823$-$3021A in the GC NGC~6624~\cite{Parent11} however implies that such putative unpulsed HE IC components may be dominated by a pulsed CR component, at least for this particular GC.

In addition to the pulsed flux, a steady flux from GCs is also expected in the VHE domain. The model of~\cite{BS07} predicted HE and VHE fluxes from GC by considering relativistic leptons escaping from the embedded MSP population and upscattering soft photons from background radiation fields via the IC process. Indeed, some MSPs may produce leptons with TeV energies due to acceleration of these particles by very large magnetospheric electric fields~\cite{Buesching08}. These leptons may be further accelerated in shocks in the GC resulting from colliding pulsar winds. 

A similar calculation of the unpulsed IC component to that of~\cite{BS07} was performed~\cite{Venter08_conf,Venter09_GC} using a particle injection spectrum calculated from first principles and which is the result of acceleration and CR losses occurring in the MSP magnetospheres. No further particle acceleration was assumed after escape from the magnetosphere, yielding predictions that should be considered as lower limits for the VHE flux band. This model predicted that 47~Tucanae and Terzan~5 may be visible for H.E.S.S., depending on the assumed model parameters, particularly $N_{\rm tot}$ and cluster magnetic field $B$. This model furthermore fixed the particle efficiency $\eta_e$ to $\sim7\%$ of the average spin-down luminosity, reducing the number of free parameters.

Recent H.E.S.S. upper limits on the TeV emission from the 47~Tucanae~\cite{Aharonian09_Tuc} implied that $N_{\rm tot}\sim30-40$ for $B\sim10\mu$G, but $N_{\rm tot}$ becoming quite larger for $B<5\,\mu$G or $B>\,30\mu$G~\cite{Venter09_GC}. Also, the \textit{Fermi} LAT HE spectrum implied that there are $N_{\rm vis}\sim50-60$ MSPs in the cluster (\cite{Venter09_GC} inferred $N_{\rm vis}\sim50$). 

H.E.S.S.\ recently detected a VHE excess in the direction of Terzan~5~\cite{Abramowski11_Ter5}, offset from the center of the GC by $4^\prime$, and having a size of $9.6^\prime \times 1.8^\prime$ (compared to the \textit{Fermi} maximum likelihood source position which is offset from the GC center by $2.4^\prime$, still within the $95\%$ source position uncertainty of $r_{95}=2.9^\prime$, and source extent of $9^\prime$). In addition, diffuse X-ray emission~\cite{Eger10}, as well as several radio structures~\cite{Clapson11} have been measured from this GC. There have also been updated measurements of Terzan~5's distance ($d=5.9\pm0.5$~kpc)~\cite{Ferraro09,Valenti07}, core radius ($r_{\rm c}=0.15^\prime$), half-mass radius ($r_{\rm hm}=0.52^\prime$), tidal radius ($r_{\rm t}=4.6^\prime$), and total luminosity ($L\sim8\times10^{5}L_\odot$)~\cite{Lanzoni10}.

In this paper, we present pulsed CR as well as unpulsed IC calculations for an ensemble of MSPs in the GC Terzan~5. Independent constraints may be derived on $N_{\rm tot}$ using both of these components, while $B$ may be constrained using synchrotron radiation (SR) and IC flux components.

\section{MODEL}
We have previously calculated the pulsed CR spectrum resulting from relativistic leptons which are accelerated in the MSP magnetospheres by large pair-starved electric fields~\cite{Venter08} as predicted by the pair-starved polar cap (PSPC) model~\cite{MH04_PS}, prior to the detection of the HE spectrum by \textit{Fermi}~\cite{Abdo10}. This calculation may now be scaled to the \textit{Fermi} data to infer $N_{\rm vis}$ (Section~\ref{sec:results}). 

We also calculate an unpulsed IC flux component, using a cumulative injection spectrum made up of electrons leaving the MSP magnetospheres after having been accelerated by these pair-starved magnetospheric electric fields, and neglecting reacceleration in the GC. Using updated structural parameters, distance, and a much larger bolometric luminosity, we calculate the radiation losses and resulting unpulsed IC fluxes shown in Fig.~\ref{fig1}, assuming Bohm diffusion and GC magnetic fields of $B = 1\,\mu$G and $B = 10\mu$G (for more details, see~\cite{Venter08_conf,Venter09_GC}). We used two radiation zones in the cluster: a core region extending from $r = 0$ up to $r_{\rm c}$, as well as a halo region extending from $r_{}\rm c$ up to $r_{\rm t}$. The steady-state particle spectrum was approximated by the product of the particle injection spectrum, and an effective times cale taking into account the IC, SR and particle escape loss time scales.

We used both bright starlight and CMB as soft photon targets in our IC calculation. The energy density of the first was assumed to be $\sim2.4\times10^4$~eV~cm$^{-3}$ and $\sim1.6\times10^3$~eV~cm$^{-3}$ for each of the regions (for a temperature $T = 4,500$~K, and due to large stellar luminosity and small core radius), while the energy density for the CMB was taken to be $\sim0.27$~eV~cm$^{-3}$ (for $T = 2.76$~K).

%
%
%

\section{RESULTS}
\label{sec:results}
Figure~\ref{fig1} shows the differential CR and IC spectra calculated for Terzan~5. Scaling the pulsed CR component~\cite{Venter09_GC} to fit the \textit{Fermi} LAT data~\cite{Abdo10} implies that the number of visible MSPs is about $N_{\rm vis} \sim 60 \pm 30$. This finding is consistent with the estimate of $180^{+90}_{-100}$ obtained by~\cite{Abdo10}, and formally presents a lower limit to $N_{\rm tot}$. However, the pair-starved model probably overpredicts the CR flux by a factor of a few, and furthermore may not be valid for all MSPs (as inferred from light curve modeling~\cite{Venter09} which imply copious pair creation in some of the MSP magnetospheres,  despite their low magnetic dipole surface fields), so that $N_{\rm vis}$ may be even larger than $\sim60$. The CR spectrum furthermore cuts off below the H.E.S.S.\ sensitivity, so that the VHE signal probably originates from IC scattering.

The VHE fluxes in Figure~\ref{fig1} include IC models from~\cite{Cheng10}, scaled predictions from~\cite{BS07}, as well as our calculated IC spectra for different values of $B$. While the prediction corresponding to $B=1$\,$\mu$G is just below the H.E.S.S.\ sensitivity, the one corresponding to $B=10$\,$\mu$G is above this detection threshold above several TeV (assuming $N_{\rm tot}=N_{\rm vis}=60$).

\begin{figure}[t]
\centering
\includegraphics[width=80mm]{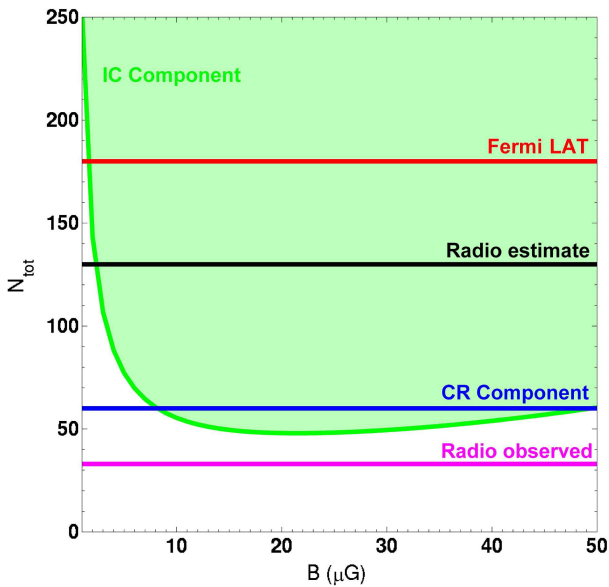}
\caption{Constraints on $N_{\rm tot}$ vs.\ GC cluster magnetic field $B$. Green line: ICS component using H.E.S.S.\ sensitivity (this work; light green area is excluded). Red and blue lines: $N_{\rm vis}$ from~\cite{Abdo10} and this work (CR component); magenta and black lines: number of radio MSPs detected and estimated~\cite{FG00}. Errors not shown.} 
\label{fig2}
\end{figure}

Figure~\ref{fig2} indicates constraints on $N_{\rm tot}$ vs.\ $B$. The green line indicates the constraint derived from our ICS calculation using H.E.S.S.\ sensitivity (the light green area is excluded). The red and blue lines are constraints on $N_{\rm vis}$ from \textit{Fermi}~\cite{Abdo10} and our CR calculation, while the magenta and black lines indicate the number of detected and estimated radio MSPs~\cite{FG00}. 

The halo size was terminated at $r_{\rm t}$, assuming that the soft photon energy density is sufficiently low outside of the GC that IC emission beyond $r_{\rm t}$ may be neglected. However, in view of the new HE and VHE data that imply that this source is extended in gamma rays, one may need to reassess this assumption. Indeed, an alternative model found~\cite{Cheng10} that most HE emission may come from a region outside of the GC core beyond a radius of 10~pc.

\section{CONCLUSION}
We have obtained pulsed and unpulsed fluxes from the GC Terzan~5, assuming CR and IC processes involving TeV leptons from a number of host MSPs. Using this model, we could constrain $N_{\rm tot}$ and $B$. Our unpulsed spectral results should be re-assessed in view of the recent H.E.S.S.\ detection of Terzan~5 in order to obtain updated constraints on these parameters. In particular, the observed spectral shape implies that reacceleration of particles may be taking place within the GC. The offset nature of the source with respect to the GC center provides a further puzzle, because if the MSPs are located within the GC core radius, the HE and VHE emission should be GC-centered. Lastly, availability of multiwavelength data on this GC may allow us to constrain the radial profile of the soft photon energy density, diffusion coefficient $k_0$ and cluster field $B$ in future.

\bigskip 
\begin{acknowledgments}
This research is based upon work supported by the South African National Research Foundation.
\end{acknowledgments}

\bigskip 

\begin{thebibliography}{99} 
\bibitem{Abdo09} Abdo, A. A. et al., Science, 2009, 325, 845
\bibitem{Abdo10} Abdo, A. A. et al., A\&A,  2010, 524, 75
\bibitem{Abramowski11_Ter5} Abramowski, A. et al., A\&A, 2011, 531, L18
\bibitem{Aharonian09_Tuc} Aharonian, F. et al. 2009, A\&A, 499, 273
\bibitem{BS07} Bednarek, W.,  Sitarek, J. 2007, MNRAS, 377, 920
\bibitem{Buesching08} B{\"u}sching, I., Venter, C., \& de Jager, O.~C., Adv. Space Res., 2008, 42, 497
\bibitem{Cheng10} Cheng, K.S. et al., ApJ,  2011, 723, 1219
\bibitem{Clapson11} Clapson, A.-C. et al., A\&A, 2011, 532, 47
\bibitem{Eger10} Eger, P., Domainko, W., \& Clapson, A.-C., A\&A, 2010, 513, A66
\bibitem{Ferraro09} Ferraro, F. R. et al., Nature,  2009, 462, 483
\bibitem{FG00} Fruchter, A. S., Goss, W. M., ApJ,  2000, 536, 865
\bibitem{Kong10} Kong, A. K. H., Hui, C. Y.,  Cheng, K.S., ApJ, 2010, 71, 36
\bibitem{Lanzoni10} Lanzoni, B. et al., ApJ, 2010, 717, 653
\bibitem{MH04_PS} Muslimov, A.~G., \& Harding, A.~K. 2004b, ApJ, 617, 471
\bibitem{Parent11} Parent, D. et al. 2011, Proc. Third \textit{Fermi} Symposium
\bibitem{Tam11} Tam, P.H.T. et al., ApJ, 2011, 729, 90
\bibitem{Valenti07} Valenti, E. et al., AJ, 2007, 133, 1287
\bibitem{Venter08_conf} Venter, C., \& de Jager, O.C. 2008, AIP Conf. Ser., 1085, 277
\bibitem{Venter08} Venter, C.,  de Jager, O.C., ApJ, 2008, 680, L125
\bibitem{Venter09_GC} Venter et al., ApJ, 2009, 696, L52
\bibitem{Venter09} Venter et al., ApJ, 2009, 707, 800
\end{thebibliography}

\end{document}